\documentclass[aps,showpacs, nofootinbib]{revtex4}
\usepackage{graphicx, graphics, color}

\newcommand {\Cbar}{\mathord{\setlength{\unitlength}{1em}
     \begin{picture}(0.6,0.7)(-0.1,0) \put(-0.1,0){\rm C}
        \thicklines \put(0.2,0.05){\line(0,1){0.55}}\end {picture}}}
\input amssym.tex
\input amssym.def

\newcommand{\be}{\begin{equation}}
\newcommand{\ee}{\end{equation}}
\newcommand{\ben}{\begin{eqnarray}}
\newcommand{\een}{\end{eqnarray}}

\newcommand{\la}{{\lambda}}

\newcommand{\cA}{{\cal A}}

\newcommand{\cO}{{\cal O}}

\newcommand{\p}{\partial}
\newcommand{\na}{\nabla}

\newcommand{\hr}{{\hat r}}

\newcommand{\Dsl}{{\slash \negthinspace \negthinspace \negthinspace \negthinspace  D}}

\newcommand{\talpha}{\tilde \alpha}


\begin{document}

\title{Euclidean Dilaton Black Hole Vortex and Dirac Fermions}

\author{{\L}ukasz Nakonieczny and Marek Rogatko}
\email{rogat@kft.umcs.lublin.pl, 
marek.rogatko@poczta.umcs.lublin.pl}
\affiliation{Institute of Physics \protect \\
Maria Curie-Sklodowska University \protect \\
20-031 Lublin, pl.~Marii Curie-Sklodowskiej 1, Poland }


\date{\today}

\begin{abstract}
We considered the behaviour of Dirac fermion modes in the background of Euclidean dilaton
black hole with an Abelian Higgs vortex passing through it.
Fermions were coupled to the fields due to the superconducting string model. The case of nonextremal
and extremal 
charged black holes in the theory with arbitrary
coupling constant between dilaton field and $U(1)$-gauge field
were considered. We 
elaborated the cases of zero and non-zero Dirac fermion modes.
One
finds the evidence that the system under consideration
can support fermion fields acting like superconducting hair on black hole
in the sence that nontrivial spinor field configuration can be carried by Euclidean spherically
symmetric charged dilaton black hole.
It was revealed that the localization of Dirac fermion modes depended on the cosmic
string winding number and the value of black hole surface gravity.

\end{abstract}

\maketitle

\section{Introduction}
In the recent years, studies of a much more realistic case than scalar fields attracted more attention.
Especially, solution of field equations describing fermions in a curved
geometry is one of the theoretical tools of investigating the underlying structure of the spacetime.
The better understanding of properties of black holes also acquires examination of the behaviour
of matter fields in the vicinity of them \cite{chandra}.  
Dirac fermions behaviour was studied in the context of Einstein-Yang-Mills background \cite{gib93}.
Fermion fields were analyzed in the near horizon limit of an extreme Kerr black hole \cite{sak04}
as well as in the extreme Reissner-Nordstr\"om (RN) case  \cite{loh84}.
It was also revealed \cite{fin00}- \cite{findir}, that 
the only black hole solution of four-spinor Einstein-dilaton-Yang-Mills equations
were those for which the spinors vanished identically outside black hole.
Dirac fields were considered in Bertotti-Robinson spacetime \cite{br1,br2}
and in the context of a cosmological solution with
a homogeneous Yang-Mills fields acting as an energy source \cite{gib94}.
\par
The late-time decay of fermion fields in the background of various kinds of black holes
is an important problem from the point of view of the uniqueness theorem for them.
The late-time behaviour of massless and massive Dirac fermion fields were widely studied
in spacetimes of static as well as stationary black holes \cite{jin04}-\cite{goz11}. 
\par
Brane models in which our Universe
is represented as $(3 + 1)$-dimensional submanifold living in higher dimensional
spacetime also  attract the attention to brane black holes. The decay of massive Dirac 
hair on a brane black hole was considered in \cite{br08}.
\par
It could happened that at the beginning our Universe underwent several phase transitions.
The mechanism of spontaneous symmetry breaking involved in the early Universe phase transitions
might produce stable topological defects like cosmic strings, monopoles and domain walls \cite{vil}.
Among them cosmic strings and cosmic string black hole
systems acquire much interest.
Assuming a distributional mass source the metric of this system was derived in \cite{ary86} (the so-called
{\it thin string limit}). 
In Ref.\cite{ach95}
the numerical and analytic evidences for the existence of an Abelian Higgs vortex on Schwarzschild black hole
were given, while in Refs. \cite{cha98}-\cite{mod98} the extensions of the aforementioned arguments to the case of
charged RN black hole and dilaton black holes were performed. 
It was also found that an analog of {\it Meissner effect} (i.e., the expulsion of the vortex fields from the black hole)
could take place. It happened that this phenomenon occurs for some range of black hole parameters \cite{bon98}.
On the contrary, extremal dilaton black holes always expel vortex Higgs fields from their interior \cite{mod99}.
The very similar
situation takes place in the case of the other topological defect, domain wall 
which can be expelled from various kinds of black holes \cite{domainwalls}.
\par
There are some cosmic strings which may become superconducting by the implementation of fermions.
They may be responsible for various {\it exotic } astrophysical phenomena. For instance,
closed superconducting loops, the so-called  vortons \cite{vor} may constitute a fraction of cold dark matter in
galactic halo, their slow quantum decays may be connected with the ultra-high energy cosmic rays (UHECRs)
\cite{nag00}. To ones dismay, it turned out that also the high-redshift gamma ray bursts could be a reasonable way
to test superconducting string model \cite{che10}. 
\par
In Ref.\cite{col91} it was revealed that Euclidean vortex solution in the spacetime of a black hole 
led to the nonperturbative exponentially decay of electric field outside the event horizon of a
Schwarzschild black hole.
It was shown \cite{dow92} that Euclidean Schwarzschild black hole could support vortex solution at the event horizon.
In Ref.\cite{mod98a} the generalization of the above problem to the case of Einstein-Maxwell-dilaton gravity was
proposed. On the other hand, in Ref.\cite{gre92} it was observed that the Dirac operator in the 
spacetime of the system composed of
Euclidean magnetic RN black hole and a vortex in the theories containing superconducting 
cosmic strings \cite{wit85} possessed zero modes. In turn, the aforementioned zero modes caused the fermion
condensate around magnetic RN black hole.
\par
The motivation of our paper was to provide some continuity with the researches 
presented in Ref.\cite{gre92} and to generalize them to the theory constituting the
modification of Einstein-Maxwell theory, the so-called dilaton gravity being the low-energy limit
of the heterotic string theory. In dilaton gravity one has to do with the non-trivial coupling of dilaton field
with the $U(1)$-gauge field. Our considerations will be valid for an arbitrary coupling constant.
In our researches we shall consider static spherically charged black hole solution in dilaton gravity
which has quite different topological structure than the one studied in \cite{gre92}.
Contrary to the researches conducted in the aforementioned reference we shall not only pay attention
to Dirac zero fermion modes but we elaborate the non-zero fermion modes in the underlying
spacetime. To our knowledge, the problem of non-zero Dirac fermion modes in the spacetime of
black hole cosmic string system has not been studied before. We 
also pay attention to the near-horizon behaviour of fermionic fields causing superconductivity
in the case of extremal charged dilaton black hole. As was mentioned before, in the spacetime
of extremal dilaton black hole one can observe the analog of {\it Meissner effect}. 
\par
The layout of our paper will be as follows.
In Sec.II we start with the discussion of the vortex itself in the background of the Euclidean
dilaton black hole being the static solution of dilaton gravity equations with 
an arbitrary $\alpha$-coupling constant. Sec.III will be devoted to the superconducting cosmic string
piercing the black hole in question. We shall elaborate the behaviour of zero Dirac fermion modes
both on non-extremal and extremal Euclidean dilaton black hole. In Sec.IV we take into account
fermion modes for the case when $k > 0$ on the same kinds of black holes. We find that Dirac 
fermion modes may be regarded as hair on the considered black holes.
In the next section we conclude our studies.

\section{Euclidean dilaton black hole /Abelian Higgs vortex system}
In the following section we shall consider
Euclidean charged dilaton black hole/ string
vortex configuration. One assumes the complete separation between the degrees
of freedom of each of the objects in question. We shall treat static charged dilaton black hole line
element as the background solution and numerically justify the existence of the vortex solution, for arbitrary
coupling constant in the considered theory.
\par  
The system under consideration will
be described by the action of the form as
\be
S = S_{1} + S_{bos},
\ee
where $S_{1}$ is the dilaton gravity action being the low-energy limit of the string action
with arbitrary coupling constant. It is provided by
\be
S_{1} = \int \sqrt{-g}~d^4 x
\left [ R - 2 ( \na \phi )^2 - e^{-2 \alpha~\phi} F_{\mu \nu}F^{\mu \nu}
 \right], 
\ee
where $F_{\alpha \beta} = 2\na_{[ \alpha}A_{\beta ]}$, $\phi$ is the
dilaton field,
$\alpha$ is a coupling constant which determines the interaction between dilaton and Abelian gauge fields.
In action $S_{1}$, the corresponding Abelian gauge field can be thought as the everyday Maxwell one.
\par
The other gauge
field is hidden in the action $S_{bos}$ and it is subject to the spontaneous symmetry
breaking. Its action implies
\be
S_{bos} = \int \sqrt{-g}~d^4 x
\left [
- (d_{\mu} \Phi)^{\dagger}d^{\mu}\Phi - \frac{1}{4} {B}_{\mu \nu}
 {B}^{\mu \nu}
- \frac{\lambda}{4} \bigg(\Phi^{\dagger}\Phi - \eta^2 \bigg)^2  \right],
\ee
where $B_{\mu \nu} = 2\na_{[\mu }B_{\nu ]}$ is the field strength associated with $B_{\mu}$-gauge field,
$\eta$ is the energy scale of symmetry breaking and $\la$ is the Higgs coupling.
The {\it covariant derivative} has the form
$d_{\mu} = \nabla_{\mu} + i~e_{R} B_{\mu}$, where $e_{R}$ is the gauge coupling constant.
\par
The line element of the general static spherically symmetric
Euclidean dilaton black hole yields
\be
ds^2 = A^2 d \tau^2 + B^2 dr^2 + C^2 (d \theta^2 +
\sin^2 \theta  d \phi^2),
\label{met}
\ee
where in order to Euclideanize the metric we set the Euclidean time as $t \rightarrow i \tau$.
For the case when $B(r)^2 = 1/A(r)^2$, 
the explicit forms of the metric coefficients are as follows: 
\ben
A^2 = \bigg( 1 - {r_{+} \over r} \bigg) \bigg(
1 - {r_{-} \over r} \bigg)^{1 - \alpha^2 \over 1 + \alpha^2}, \\
C^2 =  r^2 \bigg( 1 - {r_{-} \over r} \bigg)^{2 \alpha^2 \over 1 + \alpha^2},
\een
where $r_{+}, r_{-}$ are related to the mass and charge $Q$ of the
black hole due to the relations
\ben
2 M &=& r_{+} + {1 - \alpha^2 \over 1 + \alpha^2} r_{-}, \\
Q^2 &=& {r_{+} r_{-} \over 1 + \alpha^2} .
\een
On the other hand, the dilaton field is given as
\be
e^{2 \alpha \phi} = \bigg( 1 - {r_{-} \over r} \bigg)^{2 \alpha^2 \over 1 + \alpha^2}.
\ee
The location of the event horizon is $ r = r_{+}$, the $r = r_{-}$ is
another singularity, but one can ignore it for $r_{-} < r_{+}$.
\\
Having in mind the above charged dilaton black hole solution one can see that the structure
of the black hole in question is drastically changed due to the presence of dilaton field.
Moreover, the arbitrary $\alpha$-coupling constant is the other non-trivial element in the
studies. Recently, the numerical studies of the dynamical collapse 
of complex charged scalar field \cite{bor11} reveal that 
due to the coupling between dilaton and $U(1)$-gauge field 
the collapse leads to the Schwarzschild black hole rather than the collapse of charged
field in Einstein-Maxwell gravity. Though, when one puts coupling constant to zero
we obtain the bahaviour leading to a black hole with a Cauchy horizon. 
\par
Treating a nonlinear system coupled to gravity is a very difficult problem but it is worth
mentioning that it found in Refs.\cite{ach95}-\cite{mod99} that the self-gravitating
Nielsen-Olesen vortex can act as a long hair for various kinds of black holes.
In what follows, we refine our attention to the vortex itself and elaborate its behaviour in the background
of Euclidean dilaton black hole in the theory with arbitrary coupling constant $\alpha$.
To begin with we choose $X$ and $P$ fields provided by the expressions 
\be
\Phi(x^{i}) = \eta X(r)e^{iN  \tau \kappa},
\label{phi}
\ee
\be
B_{\mu}(x^{i}) = \frac{\kappa}{e_{R}} \left[P_{\mu}(r) - N \na_{\mu}\tau \right ] ,
\label{b}
\ee
where $\kappa = \frac{1}{2}[\partial_{r}A^2(r)]_{r = r_{+}}$ is the surface gravity
of the Euclidean dilaton black hole.
Further, we assume that $B_{\tau}$ is the only one non-vanishing coefficient of the gauge field which is subject
to the spontaneous symmetry breaking.
\par
\begin{figure}[p]
\includegraphics[scale=0.5,angle=270]{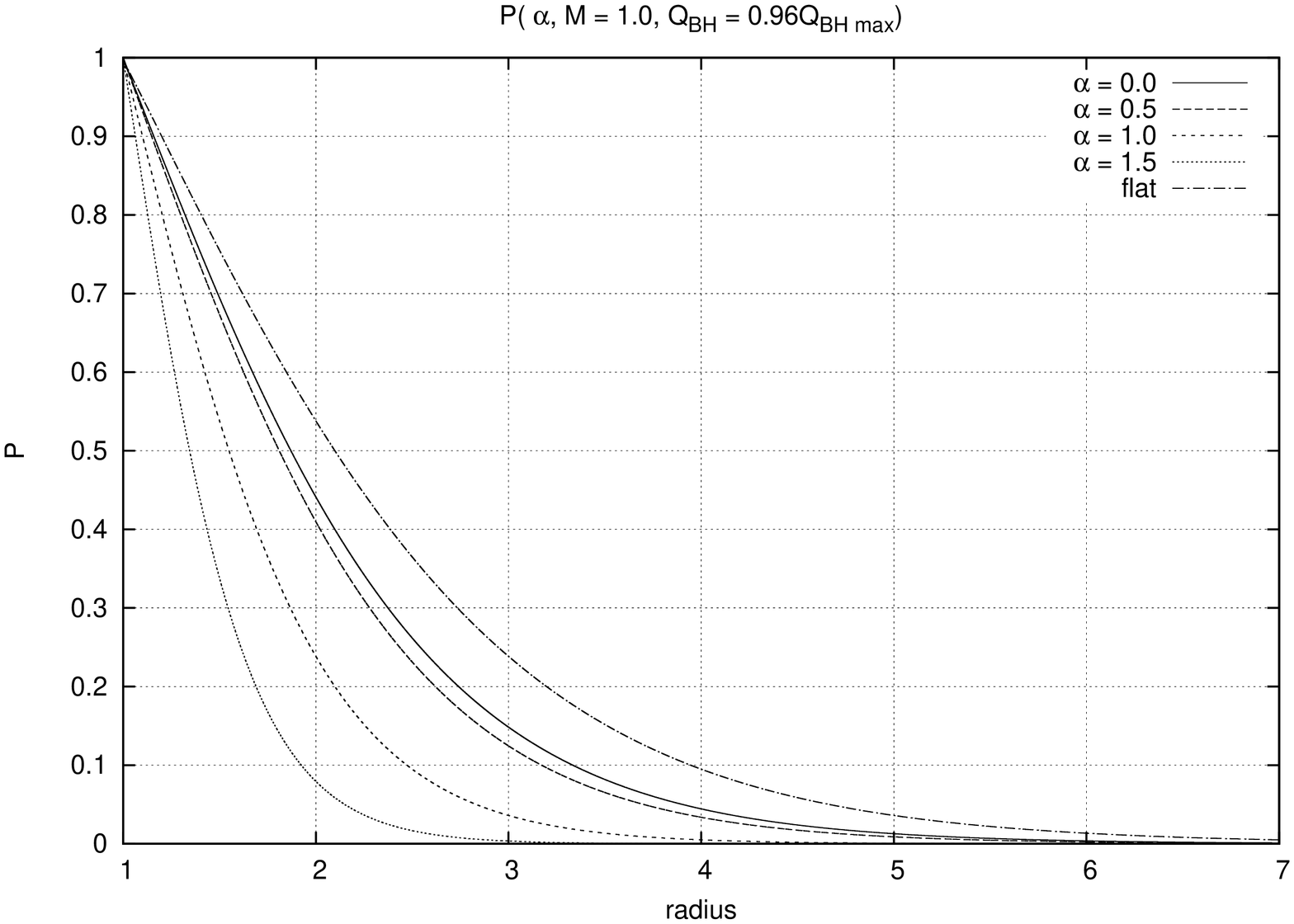}
\caption{Plot of the gauge field $P$ for the different values of the coupling constant $\alpha$.
$Q_{BH~max}$ is the charge of the extreme black hole, given by $Q_{BH~max} = \sqrt{1 + \alpha^2}M$. 
The cosmic string winding number we is equal to $1.0$.} 
\label{fig1}
\end{figure}

\begin{figure}[p]
\includegraphics[scale=0.5, angle=270]{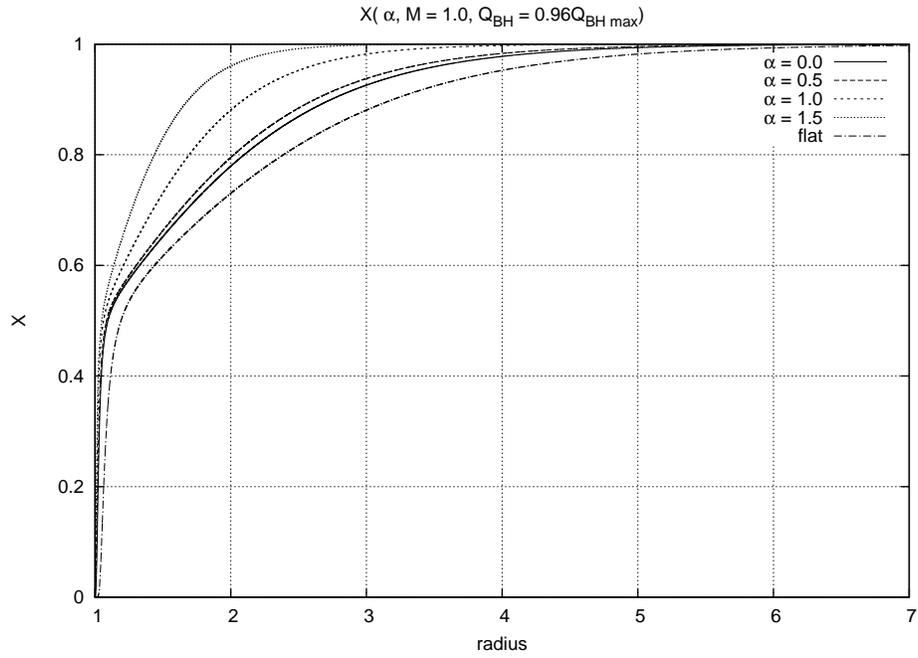}
\caption{Plot of the gauge $X$ for the different value the coupling constant $\alpha$ and for
the winding number $N = 1.0$.}
\label{fig2}
\end{figure}

\begin{figure}[p]
\includegraphics[scale=0.5, angle=270]{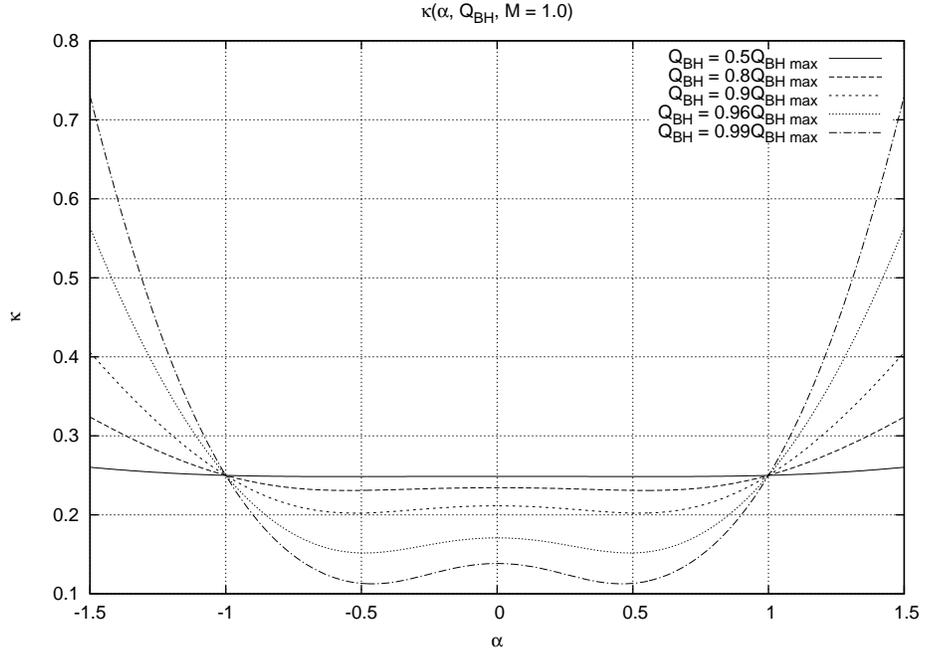}
\caption{Behavior of the surface gravity $\kappa$ for the different value of coupling constant $\alpha$.}
\label{fig3}
\end{figure}

 
Let us introduce quantities defined by
\ben
\sqrt{\lambda}\eta (M,Q_{BH},r,r_{+},r_{-}, \kappa) \equiv 
(\bar{M},\bar{Q}_{BH},\bar{r}, \bar{r}_{+},\bar{r}_{-}, \bar{\kappa}). 
\een
Taking the background solution as the
spacetime of Euclidean charged dilaton black hole, one reaches to the following equations
of motion for $X$ and $P$ fields:
\ben \label{mo1}
\frac{1}{C^2}(C^2 A^2 X_{, \bar{r}})_{, \bar{r}} &-& 
\bar{\kappa}^2 \frac{P^2 X}{A^2} -
\frac{1}{2}X(X^2 - 1) = 0, \\
\frac{1}{C^2}(C^2 P_{, \bar{r}})_{, \bar{r}} &-&
\frac{1}{\nu} \frac{X^2 P}{A^2} = 0,
\label{mo2}
\een
where we denoted $\nu = {\la \over 2e_{R}^2}$. 
The above equations can be rearranged in the forms which imply
\ben
&(1 - \frac{\bar{r}_{+}}{\bar{r}})
(1 - \frac{\bar{r}_{-}}{\bar{r}})^{\frac{1-\alpha^2}{1+\alpha^2}} 
{ d^2 \over d {\bar{r}}^2} X + 
\bigg \{ 
\frac{2}{\bar{r}} + 
\frac{2\alpha^2}{1+\alpha^2} 
\frac{\bar{r}_{-}}{\bar{r}} 
\frac{1}{\bar{r} - \bar{r}_{-}} + 
\frac{1}{\bar{r}^2}
(1 - \frac{\bar{r}_{-}}{\bar{r}})^{\frac{1-\alpha^2}{1+\alpha^2}}
[\bar{r}_{+} + 
\bar{r}_{-} 
\frac{1-\alpha^2}{1+\alpha^2}
\frac{\bar{r} - \bar{r}_{+}}{\bar{r} - \bar{r}_{-}}]  
\bigg \} 
{d \over d{\bar{r}}}X + \nonumber \\
&- \bar{\kappa}^{2} 
(\frac{\bar{r}}{\bar{r} - \bar{r}_{+}})
(\frac{\bar{r}}{\bar{r} - \bar{r}_{+}})^{\frac{1-\alpha^2}{1+\alpha^2}}
P^2 X - \frac{1}{2}X(X^2 - 1) = 0,
\label{ee1}
\een
\be
{d^2 \over d {\bar{r}}^2} P + 
\bigg \{ 
\frac{2}{\bar{r}} + 
\frac{2\alpha^2}{1+\alpha^2}
\frac{\bar{r}_{-}}{\bar{r}}
\frac{1}{\bar{r} - \bar{r}_{-}}
\bigg \}
{d \over d {\bar{r}}}P -
(\frac{\bar{r}}{\bar{r} - \bar{r}_{+}})
(\frac{\bar{r}}{\bar{r} - \bar{r}_{-}})^{\frac{1-\alpha^2}{1+\alpha^2}}
\frac{PX^2}{\nu} = 0.
\label{ee2}
\ee
We solve this set of equations numerically using the relaxation technique \cite{press}.
As in Ref.\cite{gre92} we shall work in the so-called 
supersymmetric limit when $\nu = 1$.   
Behaviour of the fields in question are depicted  in Fig.\ref{fig1} and Fig.\ref{fig2}, respectively.
For the completeness of the studies we also plotted in Fig.\ref{fig3}
the $\alpha$-dependence of the surface gravity.
It can be seen that the bounded solution at the event horizon for $P$ field 
integrated out the exponentially decaying 
at infinity while $X$ field tends to the constant value equal to $1$ at infinity.
We take the coupling constant equal to $0.0,~0.5,~1.0,~1.5$, respectively. 
On the other hand, the charge of the considered black hole
was taken by $Q_{BH} = 0.96~Q_{BHmax}$, where $Q_{BHmax} = \sqrt{1 + \alpha^2}~M$. 
In our numerical analysis
we set $r/r_{+}$ as the radial 
coordinate for the Euclidean black hole in question. For the flat spacetime we chose the ordinary $r$-coordinate.
We obtained the perfect agreement with the previous numerical studies for the case $\alpha = 0$, Ref.\cite{dow92} and
when $\alpha = 1$, Ref.\cite{mod98a}.
Our numerical investigations reveal that the larger coupling constant $\alpha$ is the quicker field $P$ tends to zero.
For the other vortex field $X$ the conclusion is similar, i.e., the larger value of $\alpha$ one considers the
quicker field $X$ tends to the constant value equal to $1$.

\section{Fermions in the Euclidean Dilaton Black Hole Background.}
In Ref.\cite{wit85} it was shown that cosmic strings can behave like superconductors and are
able to carry electric currents. In principle there are two varieties of superconducting
strings, i.e., fermionic or bosonic. In fermionic case superconductivity takes place due to
the appearance of charged Jackiw-Rossi \cite{jac81} zero modes which effectively can be regarded
as Nambu-Goldstone bosons in $1 + 1$ dimensions. They give a longitudinal component to photon
field on the cosmic string, and may be trapped in the string as massless zero modes.
On the other hand, bosonic superconductivity occurs when a charged Higgs field acquires
an expectation value in the core of the cosmic string. In this case the current in the
aforementioned object is carried by bosonic modes.
\par 
In this section we shall consider fermionic superconducting cosmic string piercing
an Euclidean charged dilaton black hole.
As was mentioned above
it turned out that it is possible for the currents to be carried by fermionic degrees of freedom confined
to the cosmic string core \cite{wit85}. 
If one takes into account the electromagnetic one, then the cosmic string will behave as superconducting.
It can be performed by extending $U(1) \times U(1)$ Lagrangian
by adding the following fermionic sector:
\be
S_{FE} = \int \sqrt{- g}~ d^4x \left [ 
\bar{\psi}\gamma^{\mu}D_{\mu}\psi +\bar{\chi}\gamma^{\mu}D_{\mu}\chi 
+ i~ \talpha~ \bigg( \Phi~ \psi^{T} C \chi - \Phi^{*}~ \bar{\psi}C \bar{\chi}^{T} \bigg) \right ],
\ee
where $\talpha$ is a coupling constant while Dirac operator satisfies the relation of the form as
\be
D_{\mu} = \nabla_{\mu} + i~ R~e_{R}B_{\mu} + i~Q~e_{q}A_{\mu}.
\label{cov}
\ee
We take that $A_{\mu} = A_{\phi} = -q_{M}\cos\theta $. The {\it covariant derivative}
for spinor fields is given by the standard relation
$\nabla_{\mu} = \partial_{\mu} + \frac{1}{2}\omega^{ab}_{\mu} \gamma_{a} \gamma_{b}$.
Dirac gamma matrices forming the chiral basis for the problem in question yield
\be
\gamma^{0} = \pmatrix{ 0&I \cr I&0 } , \qquad
\gamma^{a} = \pmatrix{ 0&i\sigma^{a} \cr -i\sigma^{a}&0 },
\label{gam}
\ee
where the Pauli matrices are provided by
\ben
\sigma^{0} = \pmatrix{ 1&0 \cr 0&1 }, \qquad
\sigma^{1} = \pmatrix{ 0&1 \cr 1&0 }, \\
\sigma^{2} = \pmatrix{ 0&-i \cr i&0 } , \qquad
\sigma^{3} = \pmatrix{ 1&0 \cr 0&-1 }. 
\een
Moreover, the charge conjugation matrix implies
\ben
C &=& \pmatrix{ -i\sigma^{2}&0 \cr 0&i\sigma^{2} }, \\
C^{\dagger} &=& C^{T} = -C.
\een
In what follows we consider the line
element of the Euclidean spherically
symmetric static black hole with a vortex passing through it.
Its metric given by (\ref{met}),  
with the line element on the $S^2$-sphere defected by the presence of the vortex.
Hence, it yields 
\be
d \Omega^2 =  C^2(r)(d\theta^2 + \tilde{b}^2 \sin \theta^2 d\phi^2),
\label{metryka-ogolna}
\ee
where $\tilde{b} = 1 - 4\mu$ is a cosmic string parameter.\\
For the above metric the curved spacetime gamma matrices are related to those given by Eq.(\ref{gam}) 
by the relations as
\ben
\gamma^{\tau} &=& A^{-1} \pmatrix{ 0&I \cr I&0 } , \qquad
\gamma^{r} =B^{-1} \pmatrix{ 0&i\sigma^{1} \cr -i\sigma^{1}&0 }, \\
\gamma^{\theta} &=& C^{-1} \pmatrix{ 0&i \sigma^{2} \cr -i\sigma^{2}&0 } , \qquad
\gamma^{\phi} =\frac{1}{C \tilde{b} \sin \theta} \pmatrix{ 0&i\sigma^{3} \cr -i\sigma^{3}&0 }.
\een
In the spacetime under consideration spinors $\psi$ and $\chi$ and their complex conjugations must be regarded
as independent fields. Following this idea,
variation of the fermion action $S_{FE}$ with respect to $\psi$ and $\chi$ fields, implies the 
following equations of motion:
\ben
\gamma^{\mu}D_{\mu}\psi - i \talpha~ \Phi^{*} C \bar{\chi}^{T} = 0, \nonumber \\
\gamma^{\mu}D_{\mu} \bar{\chi}^{\dagger} - i \talpha~ \Phi^{*} C \psi^{*} = 0,
\een
plus the analogous relations achieved by conjugations of the adequate relations in question. 
\par
With the above definitions one finds the exact form of the Dirac operator.
Accordingly, they yield the result that
\be
\Dsl = \gamma^{\mu}D_{\mu} = 
\pmatrix{ 0& D^{+} \cr D^{-}&0},
\ee
where we have denoted by $D^{+}$ and $D^{-}$ the following parts of the Dirac operator defined above
\ben
D^{+} &=& \sigma^{\tau}D_{\tau} + i\sigma^{k}~D_{k}, \\
D^{-} &=&  \sigma^{\tau}D_{\tau} - i\sigma^{j}~D_{j}.
\een
Consequently, with the remark that in Euclidean spacetime $\psi$ and $\bar{\psi}$ must be
treated as independent fields, we implicitly
choose $\psi$ and $\chi$ as left-handed, while $\bar{\psi}$ and $\bar{\chi}$ as right-handed.         
Namely, they can be brought to the forms
\ben
\psi &=& \pmatrix{ \psi_{L} \cr 0 }, \qquad \chi = \pmatrix{ \chi_{L} \cr 0 }, \\
\bar{\psi} &=& \pmatrix{0 \cr \psi_{R}}, \qquad \bar{\chi} = \pmatrix{0 \cr \chi_{R}}. 
\een
Returning to the equations of motion, they can be rewritten as
\ben 
D^{+}\psi_{R}  &+& i\talpha~ \Phi^{*}(- i \sigma^{2} \chi_{L}^{*}) = 0, \nonumber \\
D^{-}\chi_{L} &+& i\talpha~ \Phi^{*}(  i \sigma^{2} \psi_{R}^{*}) = 0.
\label{eqmotion}
\een
As in Ref.\cite{gre92} we choose the following form of $\psi$ and $\chi$ spinors:
\ben \label{pss}
\psi_{L} &=& f_{L}\xi_{-}, \qquad \chi_{R} = g_{R} \xi_{+},\\
\psi_{R} &=& f_{R} \xi_{-}, \qquad \chi_{L} = g_{L} \xi_{+},
\een
where $\xi_{\pm}$ we choose as the right and left-handed ones. They obey the relation of the form as
$
i\sigma^{2}\sigma^{3} \xi_{\pm} = \pm \xi_{\pm}
$,
which assure that fermions propagate along the cosmic string.
On the other hand, taking into account the normalization conditions $<\xi_{\pm}| \xi_{\pm}> = 1$,
one arrives at the explicit form of $\xi_{\pm}$. Namely, they may be written in the form as
\be
\xi_{+} = \frac{1}{\sqrt{2}} \pmatrix{ 1 \cr -1}, \qquad 
\xi_{-} = \frac{1}{\sqrt{2}} \pmatrix{ 1 \cr 1}.
\ee 
By virtue of the above one can readily verify that they satisfy the following:
\ben \label{act}
i \sigma^{1} \xi_{\pm} &=& \mp i \xi_{\pm},\qquad
- i \sigma^2 \xi_{\pm} = \pm \xi_{ \mp},\\
\sigma^3 \xi_{\pm} &=& \xi_{\mp}, \qquad
\sigma^0 \xi_{\pm} = \xi_{\pm}.
\een

\section{s-waves}
First of all, we shall elaborate the s-wave case. 
Behaviour of the Dirac operator acting on $S^2$-sphere with magnetic monopole and pierced by an Abelian
vortex will be crucial in this case.
Namely, on evaluating the action of the Dirac operator 
$D_{S^2}$ on $\xi_{+}$ spinor we find that it is proportional to $\xi_{-}$.
Having in mind the orthogonality condition  for $\xi_{\pm}$ one can draw  conclusion that 
the only admissible eigenvalue of $k$ is $k = 0$.
Taking into account the explicit for of the Dirac operators and Eqs.(\ref{pss})-(\ref{act}), after
a little algebra, equations of motion (\ref{eqmotion}) reduce to
\be 
A^{-1}\bigg(
i \p_{\tau} + R~{\kappa}~(P - N)\bigg)~f_{R}^{*} 
+ \bigg(
B^{-1} \p_{r} - {1 \over 2} A^{-1}B^{-1}\p_{r}A + B^{-1}C^{-1}\p_{r}C
\bigg)~
f_{R}^{*} 
+ \talpha~ \Phi~ g_{L} = 0, 
\label{s1}
\ee
and
\be
A^{-1}\bigg(
-i \p_{\tau} + R~{\kappa}~(P - N) \bigg)~g_{L}
+ \bigg(
B^{-1}\p_{r} + {1 \over 2}A^{-1}B^{-1}\p_{r}A + B^{-1}C^{-1}\p_{r}C
\bigg)~ g_{L} 
 + \talpha~\Phi^{*}~f_{R}^{*}= 0. 
\label{s2}
\ee
In what follows we assume that the time-dependence of $g_{L}$ and $f_{R}$ will be of the form
$g_{L}(\tau,~ r) = e^{- i \omega_{g} \tau}g_{L}(r) $ and $f_{R}(\tau,~ r) = e^{- i \omega_{f} \tau}f_{R}(r)$, where
$\omega_{f/g}~ \epsilon~ \Cbar$. The explicit form of the field $\Phi$ given by Eq.(\ref{phi}),
as well as Eqs.(\ref{s1})-(\ref{s2}), enables us to deduce that
$e^{- i \omega_{g} \tau} = e^{- i  N~\kappa \tau}~ e^{i  \omega_{f}^{*} \tau}$. 
Consequently, it leads to the condition
\be
\omega_{g} = N~\kappa  - \omega_{f}^{*}.
\label{euct}
\ee
As was argued in Ref.\cite{gre92} 
the action of the operators $R$ and $Q$ appearing in the definition of the {\it covariant derivative}
given by Eq.(\ref{cov}), can be defined as $R~f = \tilde{r}~ f$ and $R~g = -(\tilde{r} +1)~g$, while
$Q~f = q~f$ and $Q~g = - q~g$.  
On this account it is customary to rewrite equations of motion as follows:
\ben \label{eqq1}
A^{-1} \bigg(
&-& \omega^{*}_{f} + \tilde{r}~{\kappa}~(P - N) \bigg)
f_{R}^{*} 
+ \bigg( 
B^{-1}\partial_{r} - \frac{1}{2}A^{-1}B^{-1} {d \over dr}A + B^{-1}C^{-1}{d \over dr}C
\bigg)f_{R}^{*} 
+ m_{fer}~X g_{L} = 0,\\ \label{eqq2}
A^{-1}\bigg(
&-& {N}~{\kappa} + \omega_{f}^{*} - (\tilde {r} + 1 )~{\kappa}~ (P - N) \bigg)
g_{L} + \bigg(
B^{-1}\partial_{r} + \frac{1}{2}A^{-1}B^{-1}{d \over dr}A + B^{-1}C^{-1} {d \over dr}C
\bigg) g_{L} \\ \nonumber
 &+& m_{fer}~X f_{R}^{*}= 0, 
\een
where for brevity we have denoted $m_{fer} = \talpha \eta$.

\subsection{Nonextremal Euclidean Dilaton Black Hole}
First we elaborate the behaviour of Dirac fermions in the vicinity of the black hole event horizon.
We shall begin with considering the nonextremal case of dilaton Euclidean black hole.
Condition $A(r_{+}) = 0$ determines the outer black hole event horizon, while $C^2(r_{+}) =
{\cA \over 4 \pi}$, $\cA$ is the area of the event horizon.
Having in mind the form of the line element (\ref{met}), 
we assume the regularity of the solution at $r_{+}$. On this account, we can choose locally
cylindrical coordinates in which the aforementioned metric is regular. Namely, one has that
$\hr = {\beta_{E} /2 \pi}~A(r)$,
where $\beta_{E}$ is the period of the Euclidean time.
Moreover, regularity yields that $(A^2)'_{\mid r_{+}} = {4\pi \over \beta_{E}}.$
To proceed further, let us suppose that $f_{R}^{*}$ and $g_{L}$ are provided by the following:
\ben
f_{R}^{*} &=& x_{+}(\hr)~
exp \bigg( \int \bigg( {1 \over 2}{(P - N)\kappa \over A} - m_{fer}~X \bigg) d\hr + i \omega_{f}^{*}\tau \bigg),\\
g_{L} &=&  x_{-}(\hr)~exp \bigg( \int \bigg( {1 \over 2}{(P - N)\kappa
\over A} - m_{fer}~X \bigg) d\hr  - i \omega_{g}^{*} \tau \bigg).
\een
It helps us to rewrite Eqs. of motion (\ref{eqq1})-(\ref{eqq2}) in the form as
\be
{d \over d\hr} x_{\pm} \pm (\tilde {r} + {1 \over 2}){(P - N) \over \hr}~x_{\pm}
- {1 \over \hr}~\bigg( \omega_{\pm} \pm {1 \over 2} \bigg)
\pm m_{fer}~X~(x_{-} - x_{+}) = 0,
\label{nearh}
\ee
where we set $\omega_{+} = \omega_{f}^{*}/ \kappa$ and
$\omega_{-} = \omega_{g} / \kappa$.
\par
One can draw a conclusion that $f_{R}^{*}$ and $g_{L}$ are of order of unity
as we reach the event horizon, i.e., $\hr \rightarrow 0$. They can be regarded as
hair on the Euclidean dilaton black hole
in the sense of the nontrivial field configurations maintained by black hole event horizon.
The same observation was revealed
in the case of Euclidean RN black hole solution pierced by superconducting string \cite{gre92}.
Returning to the relations (\ref{nearh}), we observe that they
resemble equations of motion obtained for cosmic string with fermion modes in flat
Minkowski spacetime \cite{gre92}.\\
An alternative way of treating the problem is to expand 
the metric coefficients of the considered line element in the nearby of the black hole event horizon.
They will be given by the following:
\be
A^{2}(r) \simeq a(r_{+})(r - r_{+}), \qquad
B^{2}(r) \simeq  b(r_{+})(r - r_{+})^{-1}, \qquad
C^{2}(r) = C^{2}(r_{+}).
\label{coord}
\ee
Next making the change of the variables described by the relations
\be
\rho^2 = 4~b(r_{+})~(r - r_{+}), \qquad
T = {1 \over 2}\sqrt{{a(r_{+}) \over b(r_{+})}}\tau,
\ee
it can be shown that the line element of nonextremal Euclidean
black hole yields
\be
ds^2 = \rho^{2}dT^2 + d\rho^2 + C^{2}(r_{+})~d\Omega^2. 
\ee
On the other hand, the asymptotic behaviour 
of the background solutions subject 
to the equations of motion for Abelian vortex fields are provided by \cite{gre92,ach95}
\be
X \sim (r - r_{+})^{|N|/2} = \rho^{|N|}, \qquad
P \sim N - \cO (r-r_{+}) = N + \cO({\rho}^2).
\ee
Returning to the equations of motion for Dirac fermion, one can easily
verify by the above relations that they reduce to the forms as
\ben \label{casea}
{d \over d \rho}
f_{R}^{*} &-& 
\bigg( {\omega_{f}^{*} + \frac{1}{2} \over \rho} \bigg)~f_{R}^{*} + 
m_{fer}~ \rho^{|N|}~g_{L} = 0, \\
{d \over d \rho} g_{L} &+& 
\bigg({ - N~\kappa + \omega_{f}^{*} + \frac{1}{2} \over \rho}
\bigg)~g_{L} +
m_{fer}~ \rho^{|N|}~f_{R}^{*} = 0. 
\label{caseb}
\een
It will be interesting to consider the influence of the winding number $N$ on 
the behaviour of the fermion modes in question. We shall elaborate two limiting
cases of the aforementioned problem, i.e., the case when ${|N| >> 1}$ and the case
for which the winding number tends to $1$. 
\par
We shall begin with the case ${|N| >> 1}$.
The close inspection of formulae (\ref{casea}) and (\ref{caseb}) reveals that
the mass term proportional to $\rho^{|N|}$ 
can be neglected because of the fact that $\rho \rightarrow 0$ near the black hole event horizon.
On this account, one has
\ben
{d \over d \rho}f_{R}^{*} &-& 
\bigg( {\omega_{f}^{*} + {1 \over 2} \over \rho} \bigg)~ f_{R}^{*} = 0, \nonumber \\
{d \over d \rho}g_{L} &+& 
\bigg({ - N \kappa + \omega_{f}^{*} + \frac{1}{2} \over \rho} \bigg)~g_{L} = 0.
\een
It can be easily checked that the solutions of the above set of differential equations
imply
\ben
f_{R}^{*} &=& c_{1}~ \rho^{\omega_{f}^{*} + \frac{1}{2}}, \nonumber \\
g_{L} &=& c_{2}~ \rho^{N \kappa - \omega_{f}^{*} - \frac{1}{2}},
\een
where $ c_{1}$ and $c_{2}$ are constants.\\
Consistent with the requirement of the finiteness of the solutions in question
on the black hole event horizon we arrive at the condition
\be
N \kappa - \frac{1}{2} \geq Re(\omega_{f}) \geq - \frac{1}{2}.
\label{k0NN}
\ee
It could be also verified, by the direct calculations, that
$g_{L}$ and $f_{R}$ belong to the square integrable class of functions, i.e.,
$\int _{0}^{\rho}\sqrt{g}~d\rho~ |g_{L}|^2 < \infty$ and $\int _{0}^{\rho}\sqrt{g}~d\rho~ |f_{R}|^2 < \infty$.
\par
In the case when $|N| \sim 1$ equations of motion for the Dirac fermion fields
are as follows:
\ben \label{nonext}
{d \over d \rho}f_{R}^{*} &-& 
\bigg( {\omega_{f}^{*} + {1 \over 2} \over \rho} \bigg)~ f_{R}^{*} + m_{fer} \rho^{|N|}g_{L} = 0, \nonumber \\
{d \over d \rho}g_{L} &+& 
\bigg({ - \omega_{g} + \frac{1}{2} \over \rho} \bigg)~g_{L} + m_{fer} \rho^{|N|}f_{R}^{*} = 0,
\een
where we have used relation (\ref{euct}) to eliminate $N$-dependence in the second term on the left-handside
of equation (\ref{nonext}).
In order to solve Eqs.(\ref{nonext}),
we assume the following 
ansatz for the spinors $f_{R}^{*}$ and $g_{L}$:
\ben \label{n1}
f_{R}^{*} &=& \rho^{\omega_{f}^{*} + \frac{1}{2}}~\tilde{f}, \nonumber \\
g_{L} &=& \rho^{N \kappa - \omega_{f}^{*} - \frac{1}{2}}~\tilde{g}.
\een
Defining $ \beta =  |N| + N \kappa - 2\omega_{f}^{*} - 1$ and after a little of algebra we get
\ben \label{seteq}
{d \over d \rho}~\tilde{f} + m_{fer} \rho^{\beta}~\tilde{g} = 0, \nonumber \\
{d \over d \rho}~\tilde{g} + m_{fer}\rho^{2 |N| - \beta}~\tilde{f} = 0.
\een
From the first equation we find that $\tilde{g} = - 1/m_{fer}~\rho ^{-\beta}~{d \over d \rho}~\tilde{f}$.
Substituting the expression for ${d \over d \rho}~\tilde{f}$ into the second equation
of the underlying system we reach to the second order differential equation for $\tilde{f}$,
i.e., one gets 
\be
{d \over d \rho} \bigg(
\rho^{- \beta}~{d \over d \rho}~\tilde{f} \bigg) - m_{fer}^2~\rho^{2 |N| - \beta}~\tilde{f} = 0.
\label{ff}
\ee
Having in mind the explicit form of $f_{R}^{*}$ and $g_{L}$ given by (\ref{n1}), and
the condition for $\beta$, one finds that
\be
N \kappa - {1 \over 2} \geq Re(\omega_{f}) \geq - \frac{1}{2}.
\label{k0N1}
\ee
After some further calculations in which we make use of the so-called {\it Lommel's} transformation
for Bessel functions, i.e., we look for the solution of Eq.(\ref{ff}) in the form as
\be
\tilde{f} = \rho^{p}~G_{\nu}(\la ~\rho^a),
\ee
where $G_{\nu}$ is Bessel function while $p,~\la,~a$ are constants, it can be shown that
the solution of (\ref{ff}) can be written in the form as
\be
\tilde{f} = C_{1}~\rho^{1 + \beta \over 2}~I_{\nu}\bigg(
{i m \over N + 1}~\rho^{N + 1} \bigg) + C_{2}~\rho^{1 + \beta \over 2}~K_{\nu}
\bigg(
{i m \over N + 1}~\rho^{N + 1} \bigg),
\ee
where $C_{1}$ and $C_{2}$ are constants while
$\nu = 1 + \beta /2(N + 1)$. $I_{\nu}$
stands for the modified Bessel function of the first kind while $K_{\nu}$ is the Macdonald's function.
Assuming that $C_{2} = 0$ and having in mind behaviour of $I_{\nu}$
when $\rho \rightarrow 0$ one can conclude that the spinor function near the event horizon tends to the constant value
described by $I_{0}$. This condition leads us to $\beta = -1$ and to the conclusion that for 
$\omega_{f}^{*} = N(1 + \kappa)/2$ one gets the greatest value of hair near the black hole event horizon.

\subsection{Extremal Dilaton Black Hole}
In the following section we shall discuss zero fermion modes in the background of the
extremal Euclidean dilaton black hole with a vortex passing through it. In the extreme black hole
case the outer event horizon coincides with the inner one, i.e.,
$r_{-} = r_{+}$. 
In order to describe the system we use coordinates given by relation (\ref{coord})with the
only modification in $C^2(r)$ coefficient. Because of the condition for the black hole be an extremal
one, we have
\be 
C^2(r) = c(r_{+})(r - r_{+}).
\label{trans2}
\ee
By virtue of this relation the line element describing the near-horizon geometry of the
extremal Euclidean dilaton black hole yields
\be
ds^2 = \rho^2 dT^2 + d\rho^2 + \frac{c(r_{+})}{4b(r_{+})}\rho^2 d\Omega^2.
\ee
Thus, in this picture, Eqs. of motion are provided by
\ben
\label{extr}
{d \over d \rho}~f_{R}^{*} &+& 
\bigg(
- \rho^{-1}~ \omega_{f}^{*} - \frac{1}{2}\rho^{-1} + \rho^{-1}
\bigg)~f_{R}^{*} + m_{fer}\rho^{N}g_{L} = 0,  \nonumber \\
{d \over d \rho}~g_{L} &+& 
\bigg(
\rho^{-1}(- N \kappa + \omega_{f}^{*} ) 
+ \frac{1}{2}\rho^{-1} + \rho^{-1}
\bigg)~g_{L} + m_{fer}\rho^{N}f_{R}^{*} = 0.
\een
As in the previous case of nonextremal Euclidean dilaton black hole,
we first we consider ${|N| >> 1}$ case.
The mass term
proportional to $\rho^{|N|} \sim 0$ will tend to zero. Then, the solutions 
of (\ref{extr}) may be written in the form as follows:
\ben
f_{R}^{*} &=& d_{1}~ \rho^{\omega_{f}^{*} - \frac{1}{2}}, \nonumber \\
g_{L} &=& d_{2}~ \rho^{N \kappa - \omega_{f}^{*} - \frac{3}{2}},
\een
where $d_{1}$ and $d_{1}$ are constants.
The finiteness and the regularity conditions on the event horizon imply
that the following conditions are satisfied:
\be
N \kappa - \frac{3}{2} \geq Re(\omega_{f}) \geq \frac{1}{2}.
\ee
On the other hand, in the case when {$|N| \sim 1$} Eqs. of motion may be rewritten in the form as
\ben
{d \over d \rho}~f_{R}^{*} &+& 
\bigg(
{1 - \omega_{f}^{*} - {1 \over 2} \over \rho}
\bigg)~f_{R}^{*} + m_{fer}\rho^{N}g_{L} = 0, \nonumber \\
{d \over d \rho}~g_{L} &+& 
\bigg(
{{3 \over 2} - N \kappa + \omega_{f}^{*} \over \rho} 
\bigg)~g_{L} + m_{fer}\rho^{N}f_{R}^{*} = 0.
\een
Introducing the ansatz for fermion zero modes given by
\ben \label{nnn1}
f_{R}^{*} &=& \rho^{\omega_{f}^{*} - \frac{1}{2}}~\tilde{f}, \nonumber \\
g_{L} &=& ~\rho^{N \kappa - \omega_{f}^{*} - \frac{3}{2}}~\tilde{g},
\een
we obtain the same set of equations as described by relations (\ref{seteq}).
Having in mind formula (\ref{nnn1}) and the requirement 
of the finiteness of $\tilde{f}$ and $\tilde{g}$ on the black hole event horizon we reach the condition
for  $Re(\omega_{f})$. Thus, the result is provided by
\be
N \kappa  - {3 \over 2} \geq Re(\omega_{f}) \geq \frac{1}{2}.
\ee
For the completeness of our research we remark that by the same procedure as
we followed in the case of nonextremal Euclidean black hole we can cast the set of the first
order ordinary differential equations into the second order one for $\tilde{f}$, which has solution
in terms of generalized Bessel functions. The range of the parameters and conclusions
about hair on extremal Euclidean dilaton black hole are the same as in the nonextremal case.

\section{$k > 0$ Dirac Fermion Modes}
Now, we turn our attention to the case when $k > 0$. Our main aim will be
to solve Eqs.(\ref{eqmotion}) and to discuss the behaviour of Dirac fermions in the case in question.
Spinors
$\psi$ and $\chi$  
are singled out in such way to correspond with the case $k = 0$. Namely, one has that
$ i \gamma^2 \gamma^3 \tilde{\xi}_{\pm} = \pm \tilde{\xi}_{\pm} $, where
$\tilde{\xi}$ is a linear combination of $\xi$. It can be done by preferring the basis in the form
\ben
\gamma^{0} & =& \pmatrix{ 0&I \cr I&0}, \qquad
\gamma^{1} = \pmatrix{ 0&i\sigma^3 \cr -i\sigma^3}, \nonumber \\
\gamma^{2} &=& \pmatrix{ 0&i\sigma^2 \cr -i\sigma^2}, \qquad
\gamma^{3} =  \pmatrix{ 0&i\sigma^1 \cr -i\sigma^1}. 
\een
Thus, the zero mode condition can be cast into $i\sigma^2 \sigma^1 \tilde{\xi_{\pm}} = \pm \tilde{\xi_{\pm}}$,
which provides the following relation:
\be
\tilde{\xi_{+}} = \pmatrix{ 1 \cr 0}, \qquad 
\tilde{\xi_{-}} = \pmatrix{ 0 \cr 1}.
\ee
In the case under consideration,
the exact form of the Dirac operators will be
given by the expressions
\ben
D^{+} &=& A^{-1} \bigg(
\partial_{\tau} + iR~(P-N)\kappa \bigg)
\sigma^{0} +
\bigg(
B^{-1}\partial_{r} - \frac{1}{2}B^{-1}A^{-1}\partial_{r}A + B^{-1}C^{-1}\partial_{r}C \bigg)i\sigma^{3}
+ {D_{S^2} \over C}, \\
D^{-} &=& A^{-1} \bigg(
\partial_{\tau} + iR~(P-N)\kappa \bigg)
\sigma^{0} +
\bigg(
- B^{-1}\partial_{r} - \frac{1}{2}B^{-1}A^{-1}\partial_{r}A - B^{-1}C^{-1}\partial_{r}C \bigg)i\sigma^{3}
+ {D_{S^2} \over C},
\een
where the Dirac operator on $S^2$-sphere with magnetic monopole pierced by an Abelian
vortex yields
\be
D_{S^2} = i\sigma^{2} \bigg( \partial_{\theta} + \frac{1}{2}\cot \theta \bigg)
+ \sigma^{1} \bigg(
\frac{i\partial_{\phi}}{\tilde{b}\sin \theta} + Qe_{M}\tilde{b}^{-1} \cot\theta \bigg).
\ee
Moreover, the action of the $D_{S^2}$ operator on $\psi$ and $\chi$ spinors implies 
\be
iD_{S^2}\psi = k~\psi, \qquad
iD_{S^2}\chi = k~ \chi,
\ee
where $k$ are eigenvalues of the operator in question.
As far as spinors $\psi$ and $\chi$ is concerned, we set them to be linear
combinations of $\tilde{\xi}_{\pm}$
\be
\psi_{R} = \pmatrix{f_{+} \cr f_{-}}, \qquad
\chi_{L} = \pmatrix{g_{+} \cr g_{-}}.
\ee
By virtue of the above 
equations of motion are provided by the following relations:
\ben
 A^{-1}\bigg(
\partial_{\tau} &+& i R (P-N) \kappa \bigg)~f_{+} +
\bigg(
B^{-1}\partial_{r} - \frac{1}{2}A^{-1}B^{-1}\partial_{r}A + B^{-1}C^{-1}\partial_{r}C \bigg)~if_{+} 
- \frac{ik}{C}f_{+} - i\alpha \Phi^{*}g_{-}^{*} = 0,\\
A^{-1} \bigg(
\partial_{\tau} &+& i R (P-N) \kappa \bigg)~f_{-} -
\bigg(
B^{-1}\partial_{r} - \frac{1}{2}A^{-1}B^{-1}\partial_{r}A + B^{-1}C^{-1}\partial_{r}C \bigg)~if_{-} 
- \frac{ik}{C}f_{-} + i\alpha \Phi^{*}g_{+}^{*} = 0, \\
 A^{-1} \bigg(
\partial_{\tau} &+& i R (P-N) \kappa \bigg)~g_{+} +
\bigg(
-B^{-1}\partial_{r} - \frac{1}{2}A^{-1}B^{-1}\partial_{r}A - B^{-1}C^{-1}\partial_{r}C \bigg)~ig_{+} 
+ \frac{ik}{C}g_{+} + i\alpha \Phi^{*}f_{-}^{*} = 0,  \\
 A^{-1} \bigg(
\partial_{\tau} &+& i R (P-N) \kappa \bigg)~g_{-} -
\bigg(
-B^{-1}\partial_{r} - \frac{1}{2}A^{-1}B^{-1}\partial_{r}A - B^{-1}C^{-1}\partial_{r}C \bigg)~ig_{-} 
+ \frac{ik}{C}g_{-} - i\alpha \Phi^{*}f_{+}^{*} = 0. 
\een 

In order to get the regular solutions on the Euclidean black hole event horizon we take only
$f_{+}$ and $g_{-}$ components to be non-zero. Consequently, this choice reduces the number of 
equations to the following set of equations:
\ben
A^{-1}\bigg(
i\partial_{\tau} &+& \tilde{r} (P-N) \kappa \bigg)~f_{+}^{*} +
\bigg(
B^{-1}\partial_{r} - \frac{1}{2}A^{-1}B^{-1}\partial_{r}A + B^{-1}C^{-1}\partial_{r}C
- \frac{k}{C} \bigg)~f_{+}^{*} - \alpha \Phi g_{-} = 0, \\
A^{-1}\bigg(
i\partial_{\tau} &+& (\tilde{r} + 1)(P-N) \kappa \bigg)~g_{-} +
\bigg(
-B^{-1}\partial_{r} - \frac{1}{2}A^{-1}B^{-1}\partial_{r}A - B^{-1}C^{-1}\partial_{r}C 
- \frac{k}{C} \bigg)~g_{-} + \alpha \Phi^{*}f_{+}^{*} = 0. 
\een 
The dependence on the Euclidean time we be given by Eq.(\ref{euct}). First, we shall 
proceed to study equations 
of motion for the non-extremal Euclidean dilaton black hole. 
As in the preceding section when $k=0$, we study the behaviour of Dirac fermions near the 
event horizon of the Euclidean dilaton black hole. Suppose that
$f_{+}^{*}$ and $g_{-}$ yield
\ben
f_{+}^{*} = x_{+}(\hr)~exp \bigg( \int \bigg(  \frac{1}{2}\frac{(P -N)\kappa}{A}  - m_{fer}X \bigg )~d\hr + 
i \omega_{f}^{*} \tau \bigg), 
\nonumber \\
g_{-} = x_{-}(\hr)~ exp \bigg( \int \bigg(  \frac{1}{2}\frac{(P -N)\kappa}{A}  - m_{fer}X \bigg) ~ d\hr 
- i \omega_{g}\tau \bigg),
\een
where $\hr$ is defined in the same manner as in s-wave case. 
In terms the above relations the considered set of equations imply
\be
{d \over d {\hat{r}}} x_{\pm} \pm
(\tilde{r} + \frac{1}{2})\frac{(P - N)}{\hat{r}}x_{\pm} 
- \frac{1}{\hat{r}} (\omega_{\pm} \pm \frac{1}{2})x_{\pm} \mp 
\sqrt{\frac{4 \pi}{\cA}}kx_{\pm} - m_{fer}X (x_{+} + x_{-}) = 0, 
\ee
where we set $\omega_{+} = \frac{\omega_{f}^{*}}{\kappa}$ and 
$\omega_{-} = \frac{\omega_{g}}{\kappa}$.
As we take into account that near the black hole event horizon $A(r) \simeq 2 \pi \hr / \beta_{E}$
one concludes that the situation is similar to zero mode case and the spinors in question
are of order of unity. Thus, they constitute hair on the considered Euclidean black hole.\\
As in Sec.IV A, to proceed further, we expand the line element in the vicinity of the black hole event horizon. 
The corresponding
relations for spinors $g_{-}$ and $f_{+}^{*}$
may be written as
\ben \label{eqmk}
{d \over d \rho}g_{-} &+& 
\bigg(
{\omega_{f}^{*} - N \kappa \over \rho}
+ {1 \over 2 \rho} + \frac{k}{C(r_{+})}  \bigg)~g_{-} - m_{fer}\rho^{|N|}f_{+}^{*} = 0, \nonumber \\
{d \over d \rho}f_{+}^{*} &-& 
\bigg(
{\omega_{f}^{*} \over \rho} + {1 \over 2 \rho} + \frac{k}{C(r_{+})}  \bigg)~f_{+}^{*} - m_{fer}\rho^{|N|}g_{-} = 0.
\een
One elaborates two cases of the cosmic string winding number.
We shall begin with 
$|N| >> 1$. One can neglect the mass term
in relations (\ref{eqmk}) and the adequate solutions can be 
expressed in terms of
\ben
f_{+}^{*} &=& C_{1}~ \rho^{\omega_{f}^{*} + \frac{1}{2}}
e^{\frac{k}{C(r_{+})}\rho}, \\
g_{-} &=& C_{2}~
\rho^{-\omega_{f}^{*} + N \kappa - \frac{1}{2}}
e^{- \frac{k}{C(r_{+})}\rho},
\een
where we denote $C_{1}$ and $C_{2}$ as constants.
On the other hand, the requirement of
regularity on the black hole event horizon implies the following: 
\be
N \kappa - \frac{1}{2} \geq Re(\omega_{f}) \geq - \frac{1}{2}.
\label{kn0NN}
\ee
One can remark that it has  the same form as in $k = 0$ case.
\par
Proceeding to the case of $|N| \sim 1$ it turned out that
equations of motion can be simplified by setting the following ansatze for the Dirac spinors in question
\ben \label{knc}
g_{-} &=&  \rho^{-\omega_{f}^{*} + N \kappa - \frac{1}{2}}~
e^{ - \frac{k}{C(r_{+})}\rho}\tilde{g}, \nonumber \\
f_{+}^{*} &=&  \rho^{\omega_{f}^{*} + \frac{1}{2}}~
e^{\frac{k}{C(r_{+})}\rho}\tilde{f}.
\een
Consequently, one can readily verify that we arrive at the expressions
\ben \label{knz1}
{d \over d \rho}
\tilde{g} &-& m_{fer}~\rho^{2|N| - \beta}~e^{\int bd\rho}~\tilde{f} = 0, \\
{d \over d \rho}
\tilde{f} &-& m_{fer}~ \rho^{\beta}~e^{ - \int bd\rho}~\tilde{g} = 0,
\label{knz2}
\een
where we have denoted
$b = k /{C(r_{+})}$.

As in the previous considerations, extracting from equation (\ref{knz2}) $\tilde{g}$ and setting in the 
remaining expression,
enables one to obtain the second order differential equation for $\tilde{f}$. Namely, it has the form
\be
{d \over d \rho}\bigg(
e^{\int bd\rho}~\rho^{- \beta}~{d \over d \rho}\tilde{f}
\bigg) - m_{fer}^2~\rho^{2|N| - \beta}~e^{\int b d\rho}~\tilde{f} = 0.
\label{unk}
\ee
Consequently, from 
relations (\ref{knc}) we have the following:
\be
N \kappa - \frac{1}{2} \geq Re(\omega_{f}) \geq - \frac{1}{2}.
\label{kn0N1}
\ee
Summing our results for the non-zero Dirac fermion modes in the spacetime
of
nonextremal Euclidean dilaton black hole with superconducting
fermion vortex we draw a conclusion that both for $N >> 1$ and $N \sim 1$,~
$k > 0$ does not modify the intervals of admissible values of $\omega_{f}$.
We attain to the same conditions on $Re(\omega_{f})$ as in the case when $k = 0$.
Namely, relations (\ref{k0NN}) and (\ref{kn0NN}), as well as Eqs.(\ref{k0N1}) and (\ref{kn0N1})
have the same forms.
However, despite of the fact that the form of Eq.(\ref{unk}) is similar to the studied before,
it seems that it has no solution in terms of the known special functions.

\subsection{Extremal Dilaton Black Hole and Non-zero Fermion Modes}
In what follows, we shall establish some main features of the behaviour of non-zero 
Dirac fermion modes in the vicinity of the event horizon of extremal Euclidean dilaton black hole
pierced by a vortex.
Using the coordinate transformation (\ref{coord}) and (\ref{trans2}) we arrive at
the following set of the first order differential equations for $g_{-}$ and $f_{+}^{*}$:
\ben
{d \over d \rho}g_{-} &+& 
\bigg( {\omega_{f}^{*} - N \kappa + \frac{3}{2} + \tilde{k} \over \rho}
\bigg)~g_{-} - m_{fer}\rho^{|N|}f_{+}^{*} = 0, \\
{d \over d \rho}f_{+}^{*} &+& 
\bigg(
{\omega_{f}^{*} + \frac{1}{2} - \tilde{k} \over \rho}
\bigg)~f_{+}^{*} 
- m_{fer}\rho^{|N|}g_{-} = 0,
\een
where we have denoted $\tilde{k} = \sqrt{\frac{4b(r_{+})}{c(r_{+})}}k$.
\par
As in the previous sections we start with the case of
$|N| >> 1$ and neglect the mass term. Next, we take into account the ansatz for Dirac fermions
\ben
g_{-} &=& D_1~ \rho^{-\omega_{f}^{*} + N \kappa - \frac{3}{2} - \tilde{k}},
\\
f_{+}^{*} &=& D_2~ \rho^{\omega_{f}^{*} - \frac{1}{2} + \tilde{k}},
\een
where $D_{1}$ and $D_{2}$ are constants.\\
Inspection of the resultant equations of motion leads us to the conclusion that the
admissible range of $Re(\omega_{f})$ is given by
\be
N \kappa - \frac{3}{2} - \tilde{k} \geq Re(\omega_{f}) \geq \frac{1}{2} - \tilde{k}.
\ee
Next,
our attempts will be to study the case when $|N| \sim 1$. 
Without loss of generality, we assume the following ansatz:
\ben \label{kkk1}
g_{-} &=& \rho^{-\omega_{f}^{*} + N \kappa - \frac{3}{2} - \tilde{k}}
\tilde{g}, \\
f_{+}^{*} &=& \rho^{\omega_{f}^{*} - \frac{1}{2} + \tilde{k}}
~\tilde{f}. 
\label{kkk2}
\een
Hence, the underlying equations of motion simplify to the form
\ben
{d \over d \rho}\tilde{g}
&-& m_{fer}\rho^{2|N| - \tilde{ \beta}}~\tilde{f} = 0,\\
{d \over d \rho}\tilde{f} 
&-& m_{fer}\rho^{\tilde{\beta}}~\tilde{g} = 0,
\een
where now we set
\be
\tilde{\beta} = |N| + N \kappa - 1 - 2\tilde{k} - 2\omega_{f}^{*}.
\ee
The same reasoning as in the previous section lead us to the second order differential equation for
$\tilde{f}$, which yields
\be
{d \over d \rho}\bigg(
\rho^{- \tilde{\beta}}~{d \over d \rho}\tilde{f}
\bigg) - m_{fer}^2~\rho^{2|N| - \tilde{\beta}}~\tilde{f} = 0.
\label{extbh}
\ee
It can be established from Eqs.(\ref{kkk1}) and (\ref{kkk2}) that the following is satisfied:
\be
N \kappa - {3 \over 2}- \tilde{k} \geq Re(\omega_{f}) \geq \frac{1}{2} - \tilde{k}.
\ee
It worth pointing out that in both cases $N >> 1$ and for $N \sim 1$ one has diminishing of the 
admissible interval of $Re(\omega_{f})$.
On the other hand, the form of Eq.(\ref{extbh}) is the same as relation (\ref{ff}) 
so the arguments in the preceding section can be repeated leading to the conclusion that the solution of
(\ref{extbh}) may be written as a combination of modified Bessel function of the first kind plus Macdonald's
function. We also conclude that the condition for the {\it largest hair} in the near-horizon limit will be diminished by the
value of $\tilde{k}$. It will be provided by $\omega_{f}^{*} = (N(1 + \kappa) - 2\tilde{k}) /2$.


\section{Conclusions}
In our paper we have analyzed the problem of an Abelian Higgs vortex on the Euclidean dilaton black 
hole in the presence of Dirac fermion modes.
Fermions were coupled to the fields in question as in Witten's model of superconducting string \cite{wit85}.
Assuming the complete separation of the degrees of freedom of the fields in our considerations we examined behaviour
of Dirac fermion modes in the nearby of the Euclidean dilaton black hole event horizon. One studied
both the case of zero and $k > 0$ Dirac fermion modes. We took into account dilaton theory 
with arbitrary coupling constant 
$\alpha$ determining the interaction between dilaton and 
$U(1)$-gauge field. Moreover, we elaborated nonextremal 
and extremal Euclidean dilaton black hole pierced by a vortex as a background of our considerations.
\par
For zero Dirac fermion modes we obtain the different interval of the $Re(\omega_{f})$-parameter 
for different kinds of the
considered Euclidean black holes. It happened that for nonextremal one, we get 
$N \kappa - {1/2} \geq Re(\omega_{f}) \geq - {1/2}$ for the winding number
$N >> 1$ and for $N \sim 1$. On the other hand,
for extremal Euclidean dilaton black hole one has that $N \kappa  - {3/2} \geq Re(\omega_{f}) \geq {1/2}$
for both aforementioned cases of choosing $N$. Having these in mind, one can draw
a conclusion that for the non-extremal black hole the admissible interval is bigger comparing to
the interval for the extremal black hole. For both kinds of the considered black holes when $N >> 1$ one reaches
effectively to massless fermions in the near-horizon region. On the other hand,
for the winding number $N \sim 1$ the fermion mass term is not a crucial ingredient
of the underlying equations of motion. 
\par
For the case when $k > 0$, we did not observe any modification of the range of $Re(\omega_{f})$
for non-extremal black holes. On the contrary, in the case of the extremal Euclidean 
dilaton black hole non-zero vale 
of $k$ diminish the admissible intervals
of $Re(\omega_{f})$.
\par
By virtue of the above, one can readily see that localization of the Dirac fermions strongly
depends on the string winding number as well as the value of the black hole surface gravity $\kappa$.
Just, it leads to the conclusion that in some situations (the adequate value of $\kappa$ which is bounded
with the black hole mass) the presence of black hole can destroy superconductivity,
in the sence of not satisfying the inequalities for $Re(\omega_{f})$.
It turned out that
superconductivity was achieved when one had to do with small black hole.\\
To conclude one remarks that spinor fields $\bar{\psi}$ and $\chi$ can be regarded as hair on 
the Euclidean dilaton black hole in the
dilaton gravity theory with the arbitrary coupling constant $\alpha$, for both
extremal and nonextremmal black holes. Hair on the considered black holes can be understand in the sense
that there is nontrivial spinor field configuration supported by the black hole event horizon.\\
Moreover, citing the same arguments as presented in Ref.\cite{gre92} reveals that the background of a magnetically
charged Euclidean dilaton black hole pierced by a vortex is lack of the physical effect of the discrete charge.
Fermions condensate appears outside the event horizon.\\
Namely, it was stated in \cite{gre92}
that in the case of Euclidean RN black hole superconducting cosmic string system
one has to do with another $U(1)$-type global symmetry which is orthogonal
to the symmetries $U(1)_{R}$ and $U(1)_{Q}$ (orthogonal in the sence that the trace of the product of the operators
connected with those symmetries is equal to zero). But this global symmetry is anomalous
because of the fact
that the aforementioned trace of the operators products is not equal to zero.
It leads to the conclusion that this symmetry is spontaneously broken in the background
of black hole and any alledged discrete charge can be absorbed by making a $U(1)$-type transformation.
Summing it all up, in the background of magnetically charged black holes one has lack
of the physical effect of the discrete charge. On the contrary, the nearby of the black hole event horizon
is furnished with a fermion condensate violating global anomalous symmetry.\\
The key point of this phenomenon is that for zero modes there is no contribution to the partition function as 
well as to the other correlation functions which do not involve the essential number of fermions.
There is no contribution to the temperature of black hole by discrete electric charge and the screened electric charge 
does not acquire exponentially small probability outside the event horizon of the considered black hole.
However, this is not the case for $k \neq 0$.


\begin{acknowledgments}
{\L} N was supported by Human Capital Programme of European Social Fund sponsored
by European Union.
\end{acknowledgments}


\end{document}